\documentclass[acmsmall]{acmart}

\AtBeginDocument{%
  \providecommand\BibTeX{{%
    \normalfont B\kern-0.5em{\scshape i\kern-0.25em b}\kern-0.8em\TeX}}}

\setcopyright{none}
\copyrightyear{2018}
\acmYear{2018}
\acmDOI{XXXXXXX.XXXXXXX}

\acmJournal{JACM}
\acmVolume{37}
\acmNumber{4}
\acmMonth{8}

\usepackage{xspace}
\usepackage[ruled,linesnumbered]{algorithm2e}

\newcommand{\name}{Talk2Data\xspace}

\newcommand{\etal}{{\textit{et~al.}}\xspace}
\newcommand{\type}{\textit{\textbf{type}}\xspace}
\newcommand{\subspace}{\textit{\textbf{subspace}}\xspace}
\newcommand{\measure}{\textit{\textbf{measure}}\xspace}
\newcommand{\focus}{\textit{\textbf{focus}}\xspace}
\newcommand{\breakdown}{\textit{\textbf{breakdown}}\xspace}

\begin{document}


\title{\name : A Natural Language Interface for \\Exploratory Visual Analysis via Question Decomposition}


\author{Yi Guo}
\authornote{Both authors contributed equally to this research.}
\email{2010937@tongji.edu.cn}
\author{Danqing Shi}
\authornotemark[1]
\email{sdq@tongji.edu.cn}
\affiliation{%
  \institution{Tongji University}
  \city{Shanghai}
  \country{China}
}

\author{Mingjuan Guo}
\affiliation{%
  \institution{Tongji University}
  \city{Shanghai}
  \country{China}
}
\email{guomingjuan@tongji.edu.cn}

\author{Yanqiu Wu}
\affiliation{%
  \institution{Tongji University}
  \city{Shanghai}
  \country{China}
}
\email{1941923@tongji.edu.cn}


\author{Nan Cao}
\authornote{Nan Cao \& Qing Chen are the corresponding authors.}
\email{nan.cao@tongji.edu.cn}
\author{Qing Chen}
\authornotemark[2]
\email{qingchen@tongji.edu.cn}
\affiliation{%
  \institution{Tongji University}
  \city{Shanghai}
  \country{China}
}

\renewcommand{\shortauthors}{Guo and Shi, et al.}


\begin{abstract}
Through a natural language interface (NLI) for exploratory visual analysis, users can directly ``ask'' analytical questions about the given tabular data. This process greatly improves user experience and lowers the technical barriers of data analysis. Existing techniques focus on generating a visualization from a concrete question. However, complex questions, requiring multiple data queries and visualizations to answer, are frequently asked in data exploration and analysis, which cannot be easily solved with the existing techniques. To address this issue, in this paper, we introduce \name, a natural language interface for exploratory visual analysis that supports answering complex questions. It leverages an advanced deep-learning model to resolve complex questions into a series of simple questions that could gradually elaborate on the users' requirements. To present answers, we design a set of annotated and captioned visualizations to represent the answers in a form that supports interpretation and narration. We conducted an ablation study and a controlled user study to evaluate the \name's effectiveness and usefulness.
\end{abstract}




\received{20 February 2007}
\received[revised]{12 March 2009}
\received[accepted]{5 June 2009}

\maketitle





\section{Introduction}

Exploratory visual analysis is an open-ended process that involves identifying questions of interest and visualizing data-oriented answers iteratively, which is non-trivial for general users while exploring unfamiliar data~\cite{tukey1977exploratory, wongsuphasawat2015voyager, battle2019characterizing}. Natural Language Interfaces (NLI) provides an accessible approach for exploratory visual analysis, in which users can can ask questions about input data in natural language and the system reveals the corresponding data-related answers in visual forms~\cite{gao2015datatone,setlur2016eviza,narechania2020nl4dv}. 
Recent works in this topic attract great attention in both academics~\cite{sun2010articulate,gao2015datatone,setlur2016eviza,aurisano2016articulate2,narechania2020nl4dv} and industry~\cite{powerbi,tableau}. These NLIs provide a more intuitive communication approach and better user experience for data analysis. Users can directly ``tell'' the analysis system their needs or ``ask'' the system for the answers to their questions about the input data. This process greatly lowers the barriers of data analysis, but the quality of the response highly depends on the system's capability of understanding users' questions that are presented in natural language. 

\begin{figure}[!t]
\centering
\includegraphics[width=0.8\linewidth]{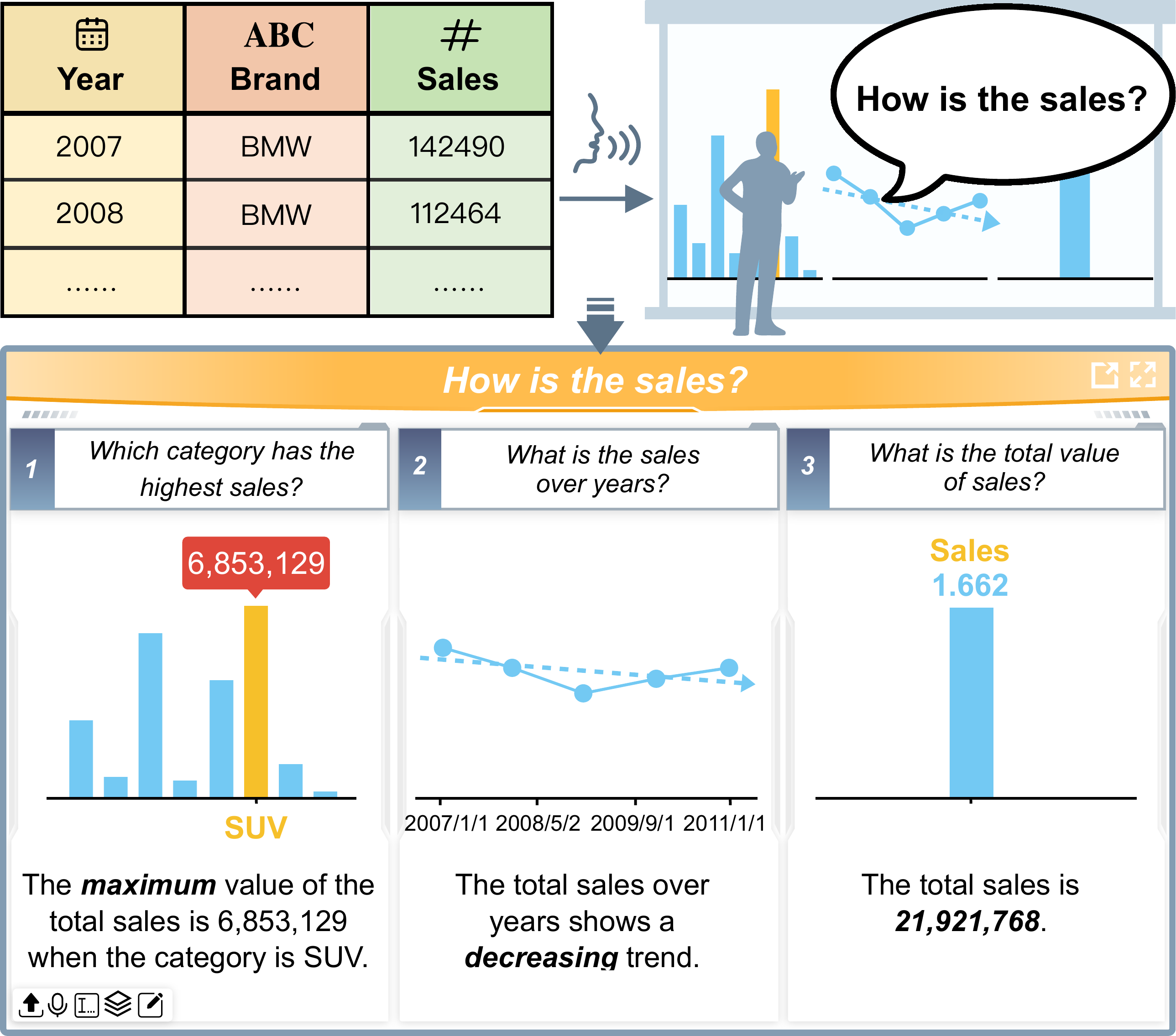}
  \caption{The user asks the system a question about car sales. The system decomposes this complex question into three simple questions and answers with a set of annotated charts.}
\label{fig:teaser}
\end{figure}


To improve the quality of response, existing techniques are designed either to guide users to provide a specific nature language query~\cite{yu2019flowsense} or untangle ambiguities in the input query~\cite{gao2015datatone} to better capture user requirements and more precisely drive the underlying data analysis for question answering. Following these ideas, most recently, Arpit~\etal introduce NL4DV~\cite{narechania2020nl4dv}, an open source python toolkit that integrates many state-of-the-art techniques~\cite{manning2014stanford,miller1995wordnet,gao2015datatone} to translate user queries into a high-level visualization grammar~\cite{satyanarayan2016vega}. The toolkit is able to map precise queries with concrete requirements to low-level analysis tasks~\cite{amar2005low}, which significantly lowers the technique barriers of building an NLI for visualization. In real-world practice, users may not only ask \textbf{simple questions} that can be directly solved by a low-level analysis task~\cite{amar2005low}. Instead, \textbf{complex questions}, such as the one shown in Fig~\ref{fig:teaser}, requiring combining multiple low-level analysis tasks are also frequently asked, which cannot be easily resolved by the existing techniques~\cite{srinivasan2017natural}.

To resolve such a complex question is difficult. Many challenges exist: first, complex questions usually cover multiple data dimensions and correspond to multiple low-level analysis tasks that are difficult for a system to differentiate.
Second, complex questions usually cannot be directly answered without mentioning the context and elaborating the details from different aspects. It is difficult to correctly extract these contextual information and details from the data merely based on a fuzzy question. Third, to better present the answers, the extracted contextual information and data details should be organized in order and visualized in a form that facilitates result narration and interpretation. 

To address the above challenges, we introduce \name, a natural language interface for exploratory visual analysis that supports natural language queries about an input spreadsheet given by both simple and complex questions. In particular, the system first employs a deep-learning based decomposition model to resolve a complex question into a series of relevant simple questions. After that, a search algorithm is introduced to explore the data space and extract meaningful data facts~\cite{srinivasan2018augmenting} that are most relevant to each simple question. These facts are finally shown in the visualizations as the parts of the answer to the input question.
We evaluate the effectiveness of the \name system via both qualitative evaluation and a controlled user study by comparing it with a baseline system developed based on NL4DV and Vega-Lite. The major contributions are as~follows:

\begin{itemize}
\item {\bf Question Decomposition.} We introduce a novel decomposition model that extends the classic sequence-to-sequence architecture~\cite{sutskever2014sequence} from four aspects: (1) A conditional vector to support the decomposition of different types of complex questions; 
(2) A decomposition layer to transform the encoding vector of the input complex question into two hidden vectors corresponding to two simple questions; 
(3) An attention mechanism~\cite{bahdanau2015neural,luong-etal-2015-effective} to enhance the relevance between the input complex question and the output simple questions.
(4) A copying mechanism~\cite{gu2016incorporating} to generalize the model to ensure it will correctly respond to the unseen datasets beyond the training corpus.




\item {\bf Visualization and System.} We designed and implemented the first natural language interface for exploratory visual analysis that supports complex questions. A set of re-designed diagrams that facilitate data narratives is also proposed and implemented in the system to represent the data facts extracted for answering the input question.
\end{itemize}
\section{Related Works}
In this section, we review the recent studies that are most relevant to our work, including natural language interfaces for visualization and question answering system.

\subsection{Natural Language Interface for Data Visualization}
Natural Language Interfaces (NLI) provide an accessible approach for data analysis, greatly lowering the requirements of user knowledge. With the goal of improving the usability of visualization NLIs, various systems have been explored both within the research community~\cite{sun2010articulate,gao2015datatone,setlur2016eviza,aurisano2016articulate2,narechania2020nl4dv,hoque2017applying,cox2001multi,wen2005optimization,srinivasan2020interweaving, liu2021advisor, luo2021natural} and industry~\cite{powerbi,tableau}. 
A common challenge for NLIs is how to precisely understand users’ intentions that are presented in a nature language (NL). In order to enhance the capability of NL interpretation, existing NLIs are designed either to guide users to provide a more concrete nature language query~\cite{cox2001multi,setlur2016eviza,yu2019flowsense} or untangle ambiguities in the input query~\cite{gao2015datatone,wen2005optimization,sun2010articulate,tableau} to better capture user' requirements and more precisely drive the underlying data analysis for question~answering. 

The initial prototype of visualization NLI~\cite{cox2001multi} does not relay on any intelligent approach to interpret user questions but create a set of supporting commands to guide users' inputs. Flowsense~\cite{yu2019flowsense} and Eviza~\cite{setlur2016eviza} depends on the pre-defined grammar to capture query patterns. When the user types a partial query and pauses, the system triggers the feature of query auto-completion to guide users' queries. However, the grammar-based methods limits the range of questions as it is impossible to cover all of the possible tasks. 

To improve the ﬂexibility in posing data-related questions while managing ambiguities in NL queries, many NLIs leverage the sophisticated NLP parsing techniques (e.g., dependencies) to understand the intuitions of queries and detect ambiguities present in the queries. In Articulate~\cite{sun2010articulate}, the translation of user’s imprecise specification is based on a NL parser imbued with machine learning algorithms that are able to make reasoned decisions automatically. 
DataTone~\cite{gao2015datatone} adopts a mixed-initiative approach to manage ambiguities in NLIs. Specifically, the system displays ambiguity widgets along with the main visualization, therefore users are allowed to switch the content in widgets to get desired alternative views. The idea in DataTone are extended to NL4DV toolkit~\cite{narechania2020nl4dv}, a python-based library released to translate user queries into a high-level visualization grammar. For the developers without experience with NLP, NL4DV can save their efforts on learning NLP knowledge when building the visualization NLIs.

The aforementioned NLIs are only able to answer the precise simple questions.
The grammatical-based methods do not support complex questions as interpreting multiple tasks in one query poses parsing difficulty~\cite{yu2019flowsense}, on the other hand, the NLP parsing techniques leverage rule-based parsers to comprehend user instructions and questions. Thus, when a question does not specifically contain keywords for analytic tasks, these interfaces cannot precisely understand user intentions. However, for the users with few knowledge about the data, it is almost impossible for them to ask precise and concrete question. In practically, the usability of these systems is under user expectations. In order to overcome these limitations, We built a novel deep-learning based model that not only can resolve a complex question into a series of relevant specific simple questions, but also improve the robustness of NL interpretation. In \name, users can ask both simple and complex questions about a table, and get well-designed diagrams that represent the facts extracted for answering the input question.
\subsection{Question Answering System}
Question Answering (QA) is a well-researched area about building systems that can answer NL questions. Advances in NLP facilitate the development of various QA systems, such as Text-Based QA~\cite{iyyer2014neural,tran2018neural,feldman2019multi,min2019multi,ding2019cognitive,perez2020unsupervised}, Knowledge-Based QA ~\cite{bao2016constraint,bao2014knowledge,berant2013semantic,saxena2020improving}, and Table-Based QA~\cite{pasupat2015compositional,yin2015neural,jauhar2016tables,yin2020tabert}.

\begin{figure*}[!t]
\setlength{\abovecaptionskip}{10pt}
\centering 
\includegraphics[width=\textwidth]{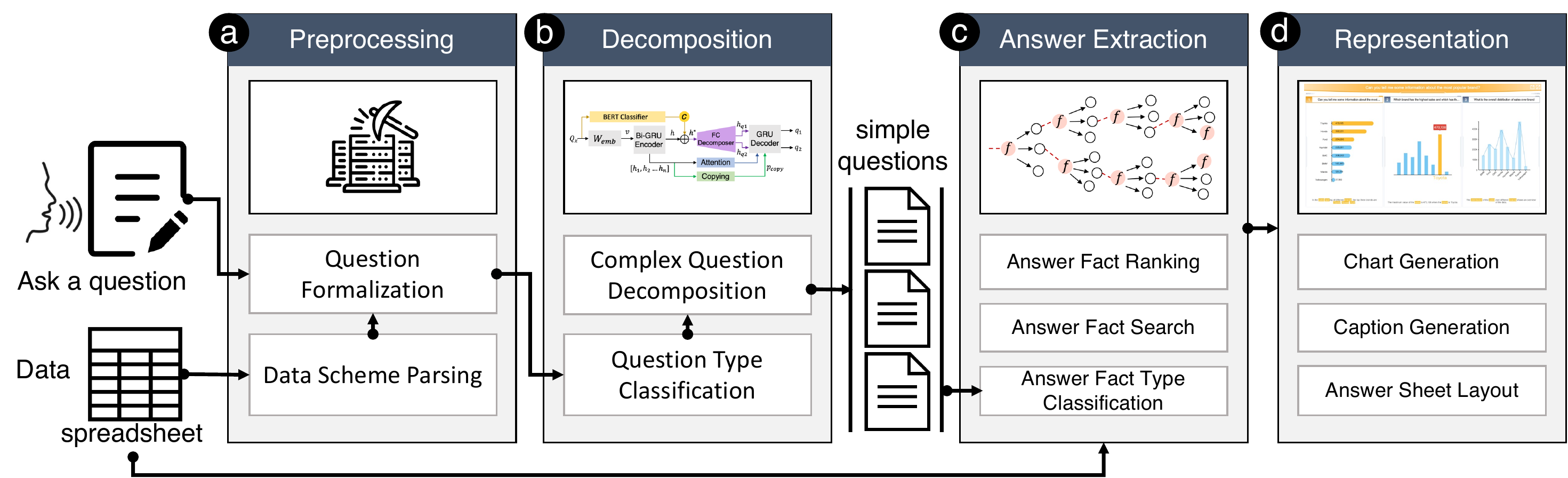}
\caption{The architecture design of \name consists of four major modules: (a) data preprocessing, (b) question decomposition, (c) answer fact extraction, and (d) visual representation.}
\label{fig:pipeline}
\end{figure*}

In this paper, our work concerns about Table-Based QA, where we are tasked to answer both complex and simple queries given a table. The existing Table-Based QA system, such as ~\cite{pasupat2015compositional,yin2015neural,neelakantan2016learning}, are designed to answer the simple questions. Given a NL query and a table, Neural Enquirer ~\cite{yin2015neural} first encodes the query and table into distributed representations, and then use a multi-layer executor to derive the answer. It can be trained using Query-Answer pairs, where the distributed representations of queries and the table are optimized together with the query execution logic in an end-to-end fashion. However, Cho \etal~\cite{cho2018adversarial} stated that only using the answer annotated dataset for training and evaluation may count ``spurious" programs, the system will accidentally lead to correct answers by using the wrong cells or
operations. Therefore, the authors propose a multi-layer sequential network with attention supervision to answer questions; it uses multiple Selective Recurrent Units to improve the interpretability of the model. Moreover, translating NL to SQL queries is a commonly used approach to answer the questions related to spreadsheets or database ~\cite{zhong2017seq2sql,xu2017sqlnet,yu2018typesql}. Seq2SQL~\cite{zhong2017seq2sql} leverage the policy-based reinforcement learning to translate the NL queries to corresponding SQL queries. SQLNET \cite{xu2017sqlnet} proposes a sequence-to-set model and a column attention mechanism to synthesize the SQL queries from NL as a slot filling problem. In the past few years, the large-scale pre-trained language models have rapidly improved the ability to understand and answer the free-form questions. The most recent work, TaBERT~\cite{yin2020tabert}, is a pre-trained language model that is built on top of BERT~\cite{devlin2018bert} to jointly learn contextual representations for queries and tables, which can be used as encoder of query and table in QA systems. 

Although efficient, the above existing Table-Based QA systems are designed to conduct information retrieval tasks, which cannot answer the questions requiring to analyzing the data. Moreover, the input questions for existing systems have to be precise and concrete. When the input question contains multiple tasks or drops the anchor words, the accuracy of system will be strongly impacted~\cite{pasupat2015compositional}. Existing works, proposed to resolving complex questions, are concentrated on text-based QA, such as ~\cite{perez2020unsupervised,min2019multi}. To our best knowledge, there is no prior study working on resolving the complex questions about a table. To fill this gap, we introduce \name, a data-oriented question and answering system that supports both simple and complex questions. In order to answer the complex questions, in \name, we adopt a novel decomposition model to resolve the complex questions into a series of simples that can be answered by data facts. The decomposition model extends the classic sequence to sequences architecture~\cite{sutskever2014sequence} and integrate attention and copy mechanism to guide the generation of each simple questions.

\section{Overview and System Design}

In this section, we describe the design requirements of the \name system, followed by an introduction of the system architecture and the problem formulation. 

\subsection{Design Requirements}
Our goal is to design and develop a data-oriented question and answering system that is able to automatically extract data facts from an input spreadsheet to answer users' complex questions about the data. To achieve the goal, a number of requirements should be fulfilled: 

\begin{enumerate}
\itemsep -1mm

\item[{\bf R1}] {\bf Elaborate complex questions in context.} The system should be able to resolve complex questions and elaborate the problem gradually from different aspects to give a comprehensive answer in context of the input data.

\item[{\bf R2}] {\bf Rank the answers.} The system should be able to rank the potential answers, i.e., data facts, in order according to their relevance to the question.

\item[{\bf R3}] {\bf Clear answers narration.} The answers to a complex question, should be visualized with narrative information such as captions and annotations and arranged in a logic order so that the users can easily read and understand them in a short time.

\item[{\bf R4}] {\bf Real-time communication and responding.} To improve the user experience, the system should be able to support real-time query and should search for the results and respond to users immediately without latency.

\end{enumerate}

\subsection{System Architecture and Formulation}
To fulfill the above requirements, as shown in Fig.~\ref{fig:pipeline}, we design \name system with four major modules: (a) the  preprocessing module, (b) the decomposition module, (c) the answer seeking module, and (d) the answer representation module. In particular, the \textit{\textbf{preprocessing module}} parses the input tabular data $X$ and the corresponding question $Q$ and combines the parsing results together as word sequence $Q_x$ in the following form to facilitate computation:
\begin{equation}
{Q_x} \leftarrow [w_1, w_2, ..., w_n, \langle N \rangle, c_{n_1}, ..., \langle T \rangle, c_{t_1}, ..., \langle C \rangle, c_{c_1}, ...]
\label{eq:question}
\end{equation}
where $w_i$ is a word / phrase in $Q$ and $c_i$ is a column in $X$ and $\langle N | T | C \rangle$ shows its corresponding data type, i.e., numerical ($N$), temporal ($T$) and categorical ($C$) respectively. 

The \textit{\textbf{decomposition module}} introduces a deep learning model by extending the classic sequence-to-sequence model to resolve a complex data-oriented question (represented by $Q_x$) into a series of relevant low-level questions that cover different aspects of the question to guide the answer seeking process ({\bf R1}):
\begin{equation}
\label{eq:decompose}
[q_{1}, q_{2}, ..., q_{m}] \leftarrow Decompose ({Q_x})
\end{equation}
where $s_i$ is a hidden vector corresponding to a low-level question that can be answered by specific data facts. The details about the decomposition algorithm is discussed in Section~\ref{sec:decomposition}.

In the \textit{\textbf{answer extraction module}}, the system searches the data space $X$ to extract data facts $f_i$ that are relevant to each of the low-level questions $q_i$ and rank them to find out the answers ({\bf R2}). The whole process is based on a parallel beam-search algorithm that guarantees the performance requirement as described in ({\bf R4}). This step can be formally presented as:

\begin{equation}
\label{eq:answer}
[f_{1}, f_{2}, ..., f_{n}] \leftarrow Extract ({q_{i}}, X)
\end{equation}
where each data fact $f_j$ is a potential answer to the question $q_{i}$. The facts are ordered based on their relevance to the question. We define the data fact $f_i$ as a 5-tuple following the definition introduced in ~\cite{shi2020calliope}, which is briefly described as follows:
\[
\begin{aligned}
    f_i &= \{type, subspace, breakdown, measure, focus\}\\
    &= \{t_i, s_i, b_i, m_i, x_i\}
\end{aligned}
\]
where \type (denoted as $t_i$) indicates the type of analysis task of the fact, whose value is one of the following cases: showing \textit{value}, \textit{difference}, \textit{proportion}, \textit{trend}, \textit{categorization}, \textit{distribution}, \textit{rank}, \textit{association}, \textit{extreme}, and \textit{outlier}; 
\subspace (denoted as $s_i$) is the data scope given by a set of filters; \breakdown (denote as $b_i$) is given by temporal or categorical data fields based on which the data items in the subspace can be divided in groups; \measure (denote as $m_i$) is a numerical data field based on which the program can retrieve a data value or compute a derived aggregated value in the subspace or each data group; \focus (denote as $x_i$) indicates a set of specific data items in the subspace that require extra attention. 

Finally, the \textbf{\textit{representation module}} organizes the data facts in order and visualize them via a set of captioned and annotated charts ({\bf R3}) that are specifically designed to help with the narration of data semantics. In the following sections, we will describe the technique details of the  decomposition, answer extraction, and representation modules.

\section{Question Decomposition}
\label{sec:decomposition}
In this section, we first introduce the two basic question decomposition methods followed by a detailed description of the decomposition algorithm and model. After that, we introduce a data corpus that we collected to train our model. Finally, we describe how we implement our algorithm.

\subsection{Question Types}

Our system is designed to resolve two types of complex questions corresponding to two ways of defining compound tasks (\textit{bottom-up} and \textit{top-down})~\cite{schulz2013design}:

\begin{itemize}
    \item \textbf{Type-I} are questions directly mentioning multiple data facts in the sentence, which can be resolved by the \textit{bottom-up} approach via enumerating and separating the simple individual questions. For example, the question ``\textit{In the year with most reviews, what is the distribution of price over different genres?}'' can be resolved into two simple questions ``\textit{Which year has the highest reviews?}'' and ``\textit{What is the overall distribution of price over genre?}".
    \item \textbf{Type-II} are questions involving multiple potential data facts due to the omission of some details, which can be resolved by \textit{top-down} approach via exploring possible individual simple questions. For instance, the question ``\textit{Which genre of book is an outlier?}'' can be resolved as simple questions such as ``\textit{Which genre of books has anomaly user ratings?}'' and ``\textit{Which genre of books has anomaly reviews?}''.
\end{itemize}

\subsection{Decomposition Algorithm}
\label{sec:decompmodel}

\begin{figure}[!tbh]
\centering 
\includegraphics[width=0.6\textwidth]{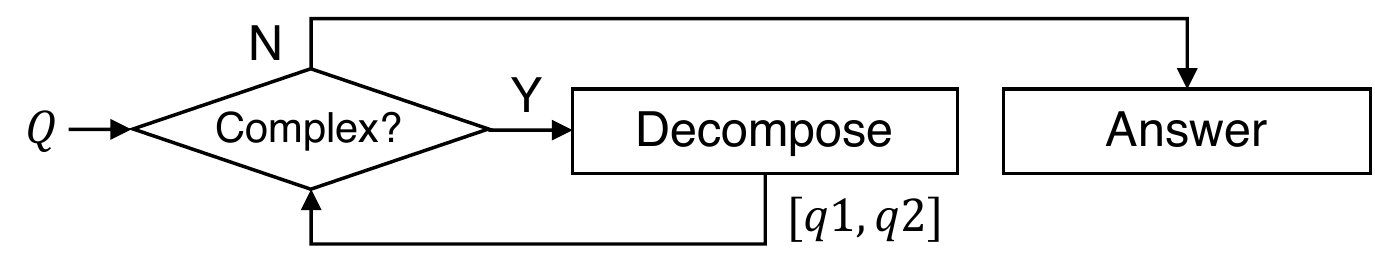}
\caption{The running pipeline of the decomposition algorithm.}
\label{fig:running}
\end{figure}

Fig.~\ref{fig:running} illustrates the running pipeline of the system's question decomposition algorithm. Given an input question $Q$ the algorithm first check if the question is a complex or simple question based on a pre-trained classifier. The simple questions will be directly send to the system's answering module, but the complex ones will be decomposed iteratively into a series of relevant simple questions via a deep decomposition model. The model resolves an formalized input complex question $Q_x$ into a set of sub-questions $(q_1, q_2, ..., q_n)$. If $q_{i}$ is a simple question, it will be directly answered, otherwise it will be decomposed again. This process runs iteratively until all the complex questions are resolved. 

\subsection{Decomposition Model}

The decomposition model is designed by extending the classic sequence-to-sequence model~\cite{sutskever2014sequence} from four aspects: (1) we pre-train a \textit{\textbf{BERT-classifier}}~\cite{devlin2018bert} to compute a conditional vector for each input question $Q_x$ to indicates the question type (i.e., {\it Type-1} or {\it  Type-II}) so a proper decomposition method could be chosen by the model; (2) we add a \textbf{\textit{decomposition layer}} in the model to transform the encoding vector $h$ of $Q_x$ into two hidden vectors ($h_{q1}, h_{q2}$) corresponding to two output questions; 
(3) we employ the \textit{\textbf{attention mechanism}}~\cite{bahdanau2015neural,luong-etal-2015-effective} in the model to ensure and enhance the relevance between the input complex question and the output questions.
(4) we integrate a \textit{\textbf{copying mechanism}}~\cite{gu2016incorporating} to generalize the model so that it could make a correct response when users ask questions about a new dataset beyond the training corpus. To make it simple, our model decomposes a complex question into exactly two sub-questions that could either be a simple question or another complex question.

\begin{figure}[!tbh]
\centering 
\includegraphics[width=0.8\textwidth]{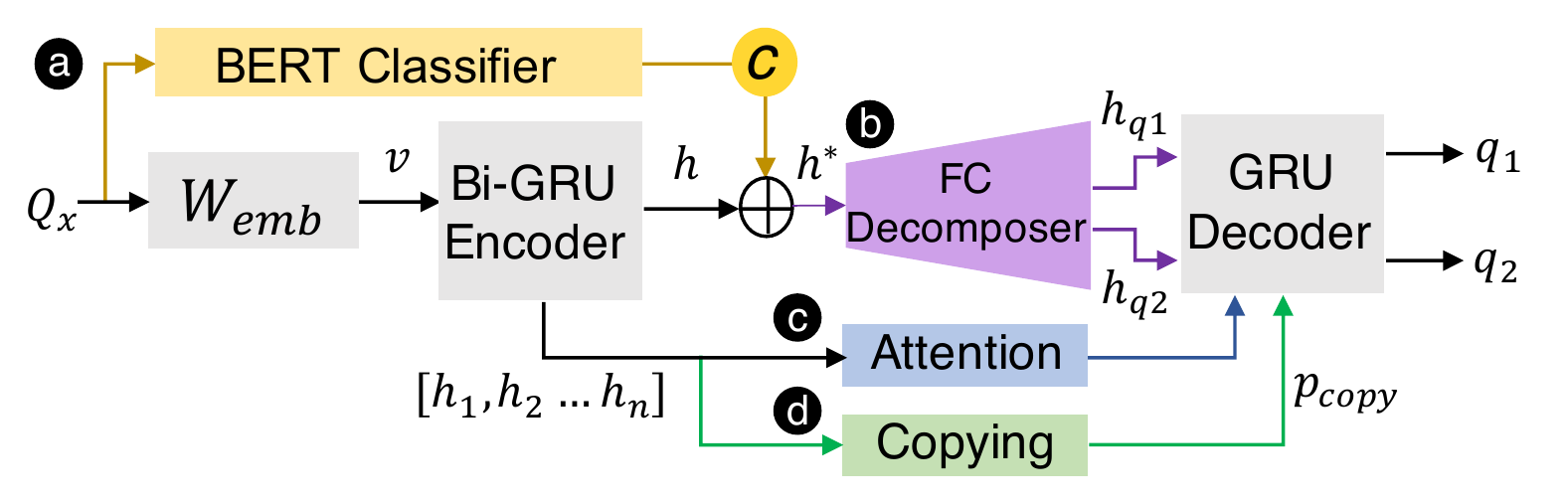}
\caption{Schematic diagrams of the decomposition model. It consists of four major improvements: (a) question classifier, (b) decomposition layer, (c) attention mechanism, and (d) copying mechanism.}
\label{fig:model}
\end{figure}

Specifically, given the formulated question $Q_x$, an embedding layer $W_{emb}$ projects each of the input words into a latent vector $v$, which is further encoded by a bidirectional GRU encoder: 
\begin{equation}
h = \textbf{encode}(v), ~~v = \textbf{embed}(Q_x)
\end{equation}
where $h$ is the vector representation of $Q_x$ that captures the semantics of the input question and the corresponding tabular data.

At the same time, as shown in Fig~\ref{fig:model}-a, a condition vector $c$ that indicates the question type is calculated by classifying $Q_x$ based on BERT~\cite{devlin2018bert}, a large scale pre-trained nature language model:
\begin{equation}
\label{eq:classification}
\begin{aligned}
c = softmax(W_{c}{u}), ~~u = \textbf{BERT}(Q_x)
\end{aligned}
\end{equation}
where $u$ is the intermediate vector encoded by BERT and $W_{c}$ is a parameter matrix to be trained. The output $c$ is a two-dimensional one-hot conditional vector that indicates the question type with $[0,1]$ indicating {\it Type-I} questions and $[1,0]$ indicating {\it Type-II} questions. With the vector, the model is able to select a proper approach to decompose the input question.

In the next, a \textbf{decomposition layer} is introduced in the model to transform $h^* = [h,c]$ into two hidden vectors $h_{q_1}$ and $h_{q_2}$ (Fig~\ref{fig:model}-b). It is implemented by a fully connected feed-forward neural network. Formally, the decomposition process is defined as follows:
\begin{equation}
\label{eq:decompose}
{h_{q_i}}=\tanh ({W_{q_i}} {h^*} + {b_{q_i}})
\end{equation}
where ${W_{q_i}}$, and ${b_{q_i}}$ are the weight matrix and bias vector to be trained.

Finally, a GRU is used to decode $h_{q_1}$ and $h_{q_1}$ to generate two sub-questions $q_{1}$ and $q_{2}$ word by word as the final output of the model:
\begin{equation}
    q_i = \textbf{decoder}(h_{q_i})
\end{equation}

In the above process, an \textit{\textbf{attention mechanism}} (Fig~\ref{fig:model}-c)~\cite{bahdanau2015neural,luong-etal-2015-effective} is incorporated to allow the decoder referencing to the relevant words in the input question when generating each word in the decomposed sub-question. It further enhances the semantic relevance between the input and output questions. At the time step $t$, the attention layer first calculates the attention weights by considering current hidden state $h_t$ in the decoder and all the hidden states of the encoder $h_{encoder} = [h_1, h_2, ..., h_n]$:

\begin{equation}
a_t = softmax(h_t^{\top} h_{encoder})
\end{equation}

The contextual vector for the input question $v_{ctx}$ is then computed as the weighted average over all the encoder states. After that, the attention layer concatenate ${v}_{ctx}$ and $h_t$ to produce a new hidden state $h^{'}_t$ for further predicting next word in the decoder:

\begin{equation}
h^{'}_t = \tanh(W_{attn}[v_{ctx};h_t]) 
\end{equation}
where $W_{attn}$ is the weight matrix to be trained.

To design a robust model, we have to consider the situation when decomposing a question about a dataset outside the scope of the training corpus. In this case, an attention mechanism is not enough as it is difficulty to predict an unseen word, e.g., the column name in the new dataset, when generating a sub-question. To address this issue, we integrated the \textit{\textbf{copying mechanism}} (Fig~\ref{fig:model}-d)~\cite{gu2016incorporating} in the model, which generalized the model by selectively copying some unseen words directly from the input (either question or data columns) when generating a sub-question. Intuitively, it estimates the probability  $p_{c} \in [0,1]$ of using a word copied from the question/data column instead of using a word generated by decoder when produces a new sub-question. Formally, $p_{c}$ is computed by combining the current hidden state $h_{t}$ in the RNN model, the contextual vector $v_{ctx}$ from the attention mechanism, and the last generated word $w_{t-1}$ together:

\begin{equation}
\label{eq:copy}
\begin{aligned}
p_{c}= sigmoid(v_{c_1}^{\top} h_{t}+ v_{c_2}^{\top} v_{ctx}+ v_{c_3}^{\top} w_{t-1})
\end{aligned}
\end{equation}
where $v_{c_i}$ is the trainable weighting vector that transform the above three vectors into a single value to predict the probability.

To encourage the output of the decomposition model as identical as possible with the target sentences in our training corpus, the model is trained by minimizing the word-level negative log likelihood loss~\cite{edunov-etal-2018-classical}:

\begin{equation}
Loss =-\sum_{i=1}^{n} \log p(t_{i} \mid t_{1}, \ldots, t_{i-1}, Q_x)
\end{equation}
where $t_i$ is the current reference word in the target sentence. Given the previous words $t_{1}, \ldots, t_{i-1}$ and the input question $Q_x$, this loss function tends to maximize the probability of the reference word $t_i$ as the prediction for current word.

\begin{table*}[!t]
    \scriptsize
    \def\arraystretch{1.2}
\begin{tabular*}{\textwidth}{cccll}
\hline
\textbf{Type} & \textbf{Method} & \textbf{\%} & \textbf{Complex Question}                                                                                                         & \textbf{Decomposed Question}                                                                                                                                    \\ \hline
                        & Comparison        & 10.0        & \begin{tabular}[c]{@{}l@{}}Please compare the list of fiction and \\ non-fiction, and rank each in order of \\ reviews for each year.\end{tabular}                   & \begin{tabular}[c]{@{}l@{}}(1) In Fiction, what is the order of reviews \\for each year? \\ (2) In Non Fiction, what is the order of reviews \\for each year?\end{tabular}                       \\ \cline{2-5} 
Type-I           & Intersection       & 13.2        & \begin{tabular}[c]{@{}l@{}}Which book is expensive and  well-regarded? .\end{tabular}                                                                                        & \begin{tabular}[c]{@{}l@{}}(1) Which book has a review higher than  average?\\                              (2) Which book has a price higher than average?\end{tabular}                           \\ \cline{2-5} 
                       & Bridging          & 27.7        & \begin{tabular}[c]{@{}l@{}}In the year with most reviews, what is the \\ distribution of price over different genre?\end{tabular} & \begin{tabular}[c]{@{}l@{}}(1) Which year has the highest/lowest reviews?\\ (2) What is the overall distribution of price \\ over genre ?\end{tabular}             \\ \hline
                       & No \textit{fact type}               & 26.5       & \begin{tabular}[c]{@{}l@{}}Show me some information about book \\ price in the different genre.\end{tabular}                      & \begin{tabular}[c]{@{}l@{}}(1) What are the differences in price between \\ each genre?\\ (2) Which genre has the highest price?\end{tabular}                      \\ \cline{2-5} 
Type-II               & No \textit{measure}            & 9.5        & \begin{tabular}[c]{@{}l@{}}Which genre of book is an outlier compare \\ with other books?\end{tabular}                            & \begin{tabular}[c]{@{}l@{}}(1) Which genre of book has an anomaly \\ user rating?\\ (2) Which genre of the book has \\ anomaly reviews?\end{tabular}               \\ \cline{2-5} 
                       & No \textit{breakdown}          & 13.1         & What is the outlier of user rating?                                                                                                 & \begin{tabular}[c]{@{}l@{}}(1) What is the outlier of user rating over different \\years?\\ (2) What is the outlier of user rating over  different \\ genre?\end{tabular} \\ \hline
                       
\end{tabular*}
\vspace{0.5em}
\caption{Examples of complex questions and the corresponding decomposed simple questions. \% shows the proportions of the questions generated by the different methods.}
\label{tab:corpustable}
\end{table*}

\paragraph{Implementation} The decomposition model was implemented in PyTorch~\cite{paszke2019pytorch}. Both encoder and decoder takes a two-layer GRU with 0.1 dropout rate for avoid over-fitting. The word embedding size and hidden size are both set to 256. The maximum length of the input sentence is 60 words. All the training parameters are initialized and updated via the Adam optimizer~\cite{kingma2014adam}, with a learning rate of 0.0001. The model was trained on a Nvidia Tesla-V100 (16GB) card.

\subsection{Training Corpus}
\label{sec:corpus}
To train our model, we prepared a new table-based question decomposition corpus (Table~\ref{tab:corpustable}) with the help of 700 English native speakers from the crowdsourcing platform\footnote{\url{https://www.prolific.co/}}. The corpus consists of 7,096 complex questions including 3,492 {\it Type-I} questions and 3,604 {\it Type-II} questions. Two simple questions were prepared as the decomposition results for each of these complex questions, i.e., 14,192 simple questions were prepared. In our corpus, we guarantee each simple question corresponds to an simple analysis task to ensure the question can be answered by at least one data fact. In general, we prepared the corpus via four steps: (1) selecting a set of meaningful data tables in various domains based on which complex questions will be prepared; (2) generating complex questions with all type of structures by a computer program; (3) manually polishing the machine-generated questions to reach the standard of natural language via the crowdsourcing platform; (4) eliminating the low-quality questions.

\textbf{Table Selection.} 
We collected 70 tabular datasets in different domains from Kaggle and Google dataset search. These datasets were further filtered based on three criteria: (1) containing meaningful data column headers; (2) having sufficient data columns and diverse column types to support all types of questions; and (3) containing informative data insights. As a result, 26 data tables were selected. Each of them has a meaningful column header and contains at least one numerical, one temporal, and one categorical field. 
All of them were tested by the online auto-insights platform~\cite{shi2020calliope} to make sure meaningful data insights could be discovered from the data. We also make the size of the data diverse, the number of rows ranges from 26 to 86,454 (mean 7,067); the number of columns ranges from 4 to 16 (mean 9). 

\textbf{Question Generation.}
To generate a complex question, we created a set of random facts and select the insightful ones based on the methods introduced in~\cite{shi2020calliope}. After that, we enumerated the facts to generate meaningful fact combinations that are potential answers to a complex question based on the methods introduced in ~\cite{schulz2013design,min2019multi}. Finally, the fact combinations are translated into a complex question based on over 200 manually prepared question templates. We traversed all 26 selected data tables and generated 5,500 {\it Type-I} questions and 7,500 {\it Type-II} questions, respectively.

Specifically, to create the {\it Type-I} questions, we selected data facts via three question reasoning methods, i.e., comparison, intersection, and bridging, as introduced in~\cite{min2019multi}. In particular, the ``comparison" type of questions compare facts within different data scopes based on the same measurement.
The ``intersection" type of questions seek for data elements that satisfy a number of conditions specified by different data facts.
Finally, the ``bridging" type of questions asks for a data fact that satisfies a prior condition specified by another. 

To generate {\it Type-II} questions, we consider three forms of complex questions: (1) the questions without mentioning an analysis task (i.e., no fact type)~;
(2) the questions without mentioning the aspect to be estimated (i.e., no measure)~;
(3) the questions without mentioning data divisions (i.e., breakdown methods).
Obviously these questions do not have a unique answer. 
Therefore, we choose all the facts that potentially answers such a complex question to help generate questions in these three forms.

\textbf{Question Rephrasing.} 
To produce high-quality natural language questions, we employed a group of native English speakers to rephrase and polish the generated questions manually. An online system was developed to split the job by randomly allocating 50 machine-generated questions to each participant. They were asked to fix the grammar errors and polish the questions into a natural language representation without changing their original meanings. 
Finally, 700 native English speakers were involved in our job and 35,000 rephrased questions were collected with an average cost of 0.12 USD per question.

\textbf{Question Validation.}
To ensure a high-quality corpus, we validated the rephrased questions through a strict process. First, all the empty and short (less than 3 words) submissions are eliminated from the corpus. After that, from each of the 50 questions processed by a participant, we manually review a 10\% question sample. Any problem found in the sample will result in an immediate rejection of all the questions rephrased by the same participant. 

Finally, we checked both the semantic $S_s(\cdot)$ and text $S_t(\cdot)$ similarities between the machine-generated $q_m$ and the rephrased $q_r$ questions to eliminate the questions that simply copy the original sentence or greatly alter the original meaning based on the following metric: 
\begin{equation}
S(q_{m}, q_{r}) = S_{s}(q_{m}, q_{r}) - S_{t}(q_{m}, q_{r})
\end{equation}
where we employ sentence-BERT~\cite{reimers2019sentence} to project a machine-generated question $q_{m}$ and rephrased questions $q_{r}$ into the same vector space and estimate their $S_s(\cdot)$ based on the cosine-similarity between the corresponding vectors. We computed the text similarity based on the Levenshtein distance $D(q_m, q_r)$, which directly estimates the word differences between two sentences that is formally defined as:
\begin{equation}
S_t(q_m,q_r) = 1 - \frac{D(q_m,q_r)}{max(|q_m|, |q_r|)}
\end{equation}
Intuitively, a positive $S(s_m, s_r)$ score indicates $s_r$ and $s_m$ share the similar semantics but are different in the text representation, i.e., $s_r$ is a high-quality rephrasing of $s_m$. In opposite, a negative $S(\cdot)$ score indicates the two questions share a similar textual representation but have different meanings, which should be eliminated.

\section{Answer Extraction}
\label{sec:answering}

In \name, the answer extraction module searches the data space $X$ to extract data facts $f_i = \{type, subspace, breakdown, measure, focus\}$ that are relevant to each of the simple questions $q_i$ and rank them to find out the answers. To balance performance and efficiency, we employ the BEAM search algorithm, which reduces computation cost when searching in large space, to retrieve the relevant facts $f_i$ for simple questions $q_i$. As shown in Fig~\ref{fig:model}, the answer extraction module consists of three parts, including (a) fact classification, (b) fact search, and (c) fact ranking. 

\begin{figure}[!htb]
\centering 
\includegraphics[width=0.8\textwidth]{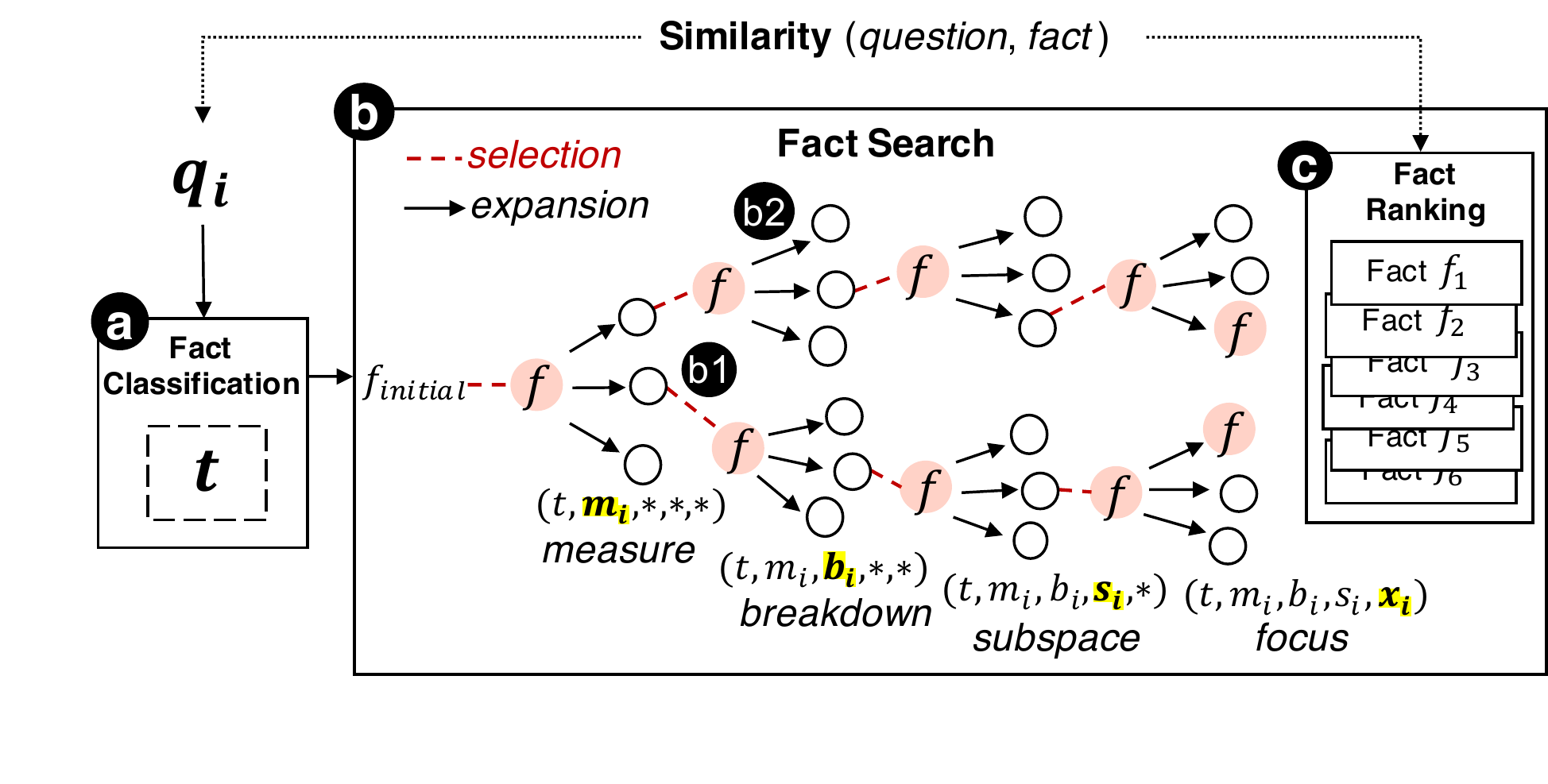}
\caption{The pipeline of answer fact extraction. It consists of three components: (a) type classification, (b) searching, and (c) ranking.}
\label{fig:model}
\end{figure}

\subsection{Fact Classification} 
The first thing to retrieve an answer fact $f_i$ to a simple question $q_i$ is to figure out the analysis tasks that are mentioned or implied in a question, so that the fact type could be determined and the rest fields in the fact could be explored guiding by the fact type. To this end, we pre-trained a
\textit{\textbf{BERT-classifer}} to indicate the \textit{type} of fact $f_i$ given a simple question $q_i$. We fine-tuned the BERT model by combining an additional classification layer and trained it on our corpus to classify the input simple question:


\begin{equation}
\label{eq:classification}
\begin{aligned}
t = softmax(W_{t}{u}), ~~u = BERT(q_i)
\end{aligned}
\end{equation}
where $u$ is the intermediate vector encoded by the BERT model and $W_{t}$ is the trainable matrix in the classification layer. The output $t$ is a ten-dimensional one-hot vector that represents the \textit{type} of answer fact $f_i$ to the input simple question $q_i$.


\subsection{Fact Search} 
Given a preferred fact type ($t_i$), we employ the beam search algorithm~\cite{ow1988filtered} to determine the rest of fact fields, i.e., subspace ($s_i$), breakdown ($b_i$) , measure ($m_i$) and focus ($x_i$) by searching through the entire data space $X$. In particular, we use semantic similarities between the resulting facts $f_i$ and the input question $q_i$ as the heuristic function to guide the searching process with the goal of retrieving facts that are most relevant to the question. As shown in Fig~\ref{fig:model} and summarized in Algorithm~\ref{alg:search}, the algorithm explores the data space $X$ to examine a number of candidate choices for each fact field step by step via a search tree $T$. In each searching step, the algorithm chooses a candidate value (e.g, a data column) for the corresponding fact field that makes the fact having the highest reward. Finally, the fact is determined by a path from the root to a leaf in the search tree. 



The search tree $T$ is gradually generated through a searching process as described in Algorithm~\ref{alg:search}. In particular, the algorithm takes a simple question $q_i$, a fact type $t$, a tabular data $X$, and a beam width $k$ (the number of selected facts at each round) as the input and automatically generates a set of data facts $\mathcal{F}$ that are most relevant to the question. At the beginning, the type $t$ initializes a data fact $f_{initial}$ to confirm the rest fields in the fact, and use $f_{initial}$ as the root of the $T$ (line 1, Fig~\ref{fig:model}-b). In next, the algorithm generates data facts by iteratively searching the rest fields in facts via three major steps: \textbf{\textit{selection}}, \textbf{\textit{expansion}}, \textbf{\textit{ranking}}. The first step selects $T_{top}$ that contains the highest $k$ facts ranked in $T$, from which the next expansion step will be performed (\textit{line 3}, Fig~\ref{fig:model}-b1). The second step expands the $T$ by creating a set of data facts. (\textit{line 4 - 8}, Fig~\ref{fig:model}-b2). The new data facts are generated by filling the field $p_i$ in $f_i$ with candidate values from $X$. The third step ranks the new facts in $T$ based on the semantic similarities between $f_i$ and question $q_i$ (\textit{line 10}). The top $k$ facts ranked in $T$ are identified as the most relevant facts for the question $q_i$  (\textit{line 12}).

\begin{algorithm}[!htb]
\label{alg:search}
\SetAlgoLined
\SetKwInOut{Input}{Input}
\SetKwInOut{Output}{Output}
\SetKw{KwBy}{by}
\Input{$q_i$, $t$, $X$, $k$}
\Output{$\mathcal{F}=[f_{1},f_{2},...,f_{k}]$} 
$ f_{initial} \leftarrow (t,*,*,*,*) $ ; 
$ T  \leftarrow [f_{initial}]$;
$ \mathcal{F}  \leftarrow []$\;

\tcp{{\small Gradually determine the rest of the fact fields, following the order of measure, breakdown, subspace, focus}}\
\For{$ p_i \in \{measure, breakdown ,subspace, focus\}$}{
           \tcc{{\small 1.selection}}\
         $T_{top} \leftarrow select(T|k)$\;
        \tcc{{\small\textcolor{blue}{2.expansion}}}
        \For{$ f_i \in T_{top} $}
        {
           \tcp{{\small If the fact field $p_i$ is not required in $f_i$, skip the $p_i$ }}\
            \If{$p_i$ is not required in $f_i$}
            {
                continue;
            }
           \tcp{{\small Expand the $T$ by creating a set of data facts, each fact is generated by filling the field $p_i$ in $f_i$ with a candidate value from $X$ }}\
         $T \leftarrow expand(f_i |p_i,X)$\;
         
        }
             \tcc{{3.ranking}}\
        $T \leftarrow rank(T | q_i)$\;
    }
   \tcp{{\small  After all the fact fields are filled, the top $k$ facts ranked in $T$ are identified as the most relevant data facts for the question $q_i$ }}\
    $\mathcal{F} \leftarrow select(T|k)$ \;
\Return $\mathcal{F}$\;
\caption{Answer Fact Searching}
\label{alg:search}
\end{algorithm}

\subsection{Fact Ranking} 
To retrieve a set of facts $\mathcal{F}$ that are most relevant to the input simple question $q_i$, in each round of expansion, the facts $f_i$ in $T$ are ranked by their semantic similarity between $f_i$ and $q_i$. As facts $f_i$ are in the form of 5-tuple, we first use hand-written templates to transform facts into machine-generated questions. If the facts $f_i$ are not complete, questions will be generated based on existing fields. Then we employ the Sentence-BERT~\cite{reimers2019sentence} to project machine-generated questions and the input question into the same vector space and use the cosine-similarity between corresponding vectors to estimate their semantic similarity.

\section{User Interface and Visualization}
\label{sec:visualization}
In this section, we introduce the representation module of the \name system. We demonstrate the design of the system's user interface and the corresponding interactions. After that, we introduce how the data facts are visually represented by a library of annotated charts and arranged in order to answer the input question.

\subsection{User Interface and Interactions}

\begin{figure}[!t]
\centering 
\includegraphics[width=0.9\textwidth]{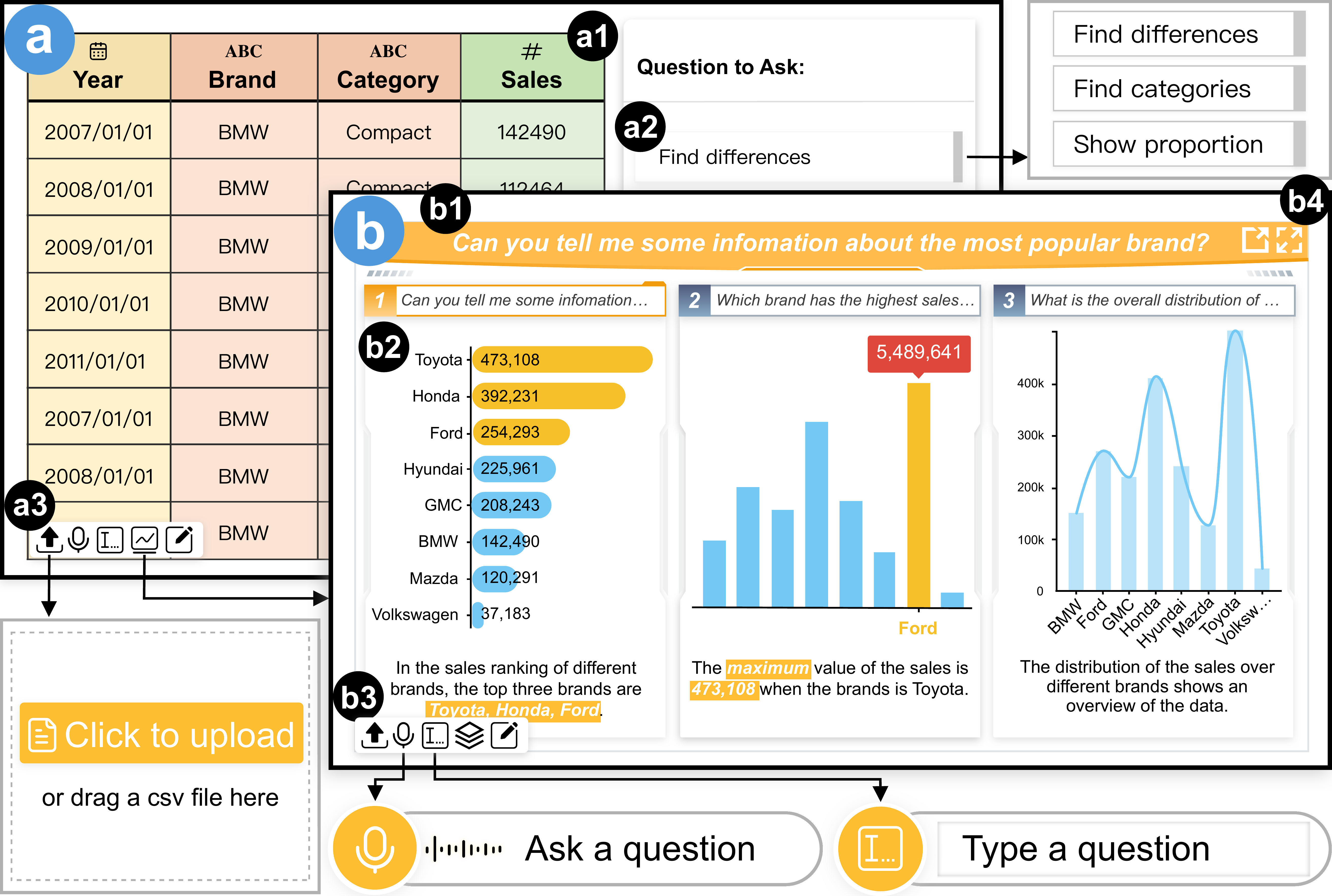}
\caption{The user interface of the \name system. 
}
\label{fig:table}
\end{figure}

The user interface of the \name system consists of two views: (1) the \textit{data view} (Fig.~\ref{fig:table}(a)) and (2) the \textit{answer view} (Fig.~\ref{fig:table}(b)). In particular, the data view is designed to illustrate the raw data to users so that they could initiate a question. In particular, the data is shown in a data table (Fig.~\ref{fig:table}(a1)) whose columns are colored by the corresponding data types. A question panel (Fig.~\ref{fig:table}(a2)) is also provided in the view to display potential questions that could be asked about the data. When a data column is selected, the question list will be updated accordingly to show questions that are only relevant to the selected column. The \textit{answer view} (Fig.~\ref{fig:table}(b)) represents data facts that answer the question via a library of annotated charts, the charts are arranged in order to facilitate the interpretation and narration of the answers. In particular, in this view, user's question is represented as the title of view (Fig.~\ref{fig:table}(b1)), and decomposed questions are shown as sub-titles in each section that are answered by data facts (Fig.~\ref{fig:table}(b2)). Each data fact is visualized by an annotated chart with a narrative caption. 

\begin{figure*}[!t]
\setlength{\abovecaptionskip}{10pt}
\centering 
\includegraphics[width=0.9\textwidth]{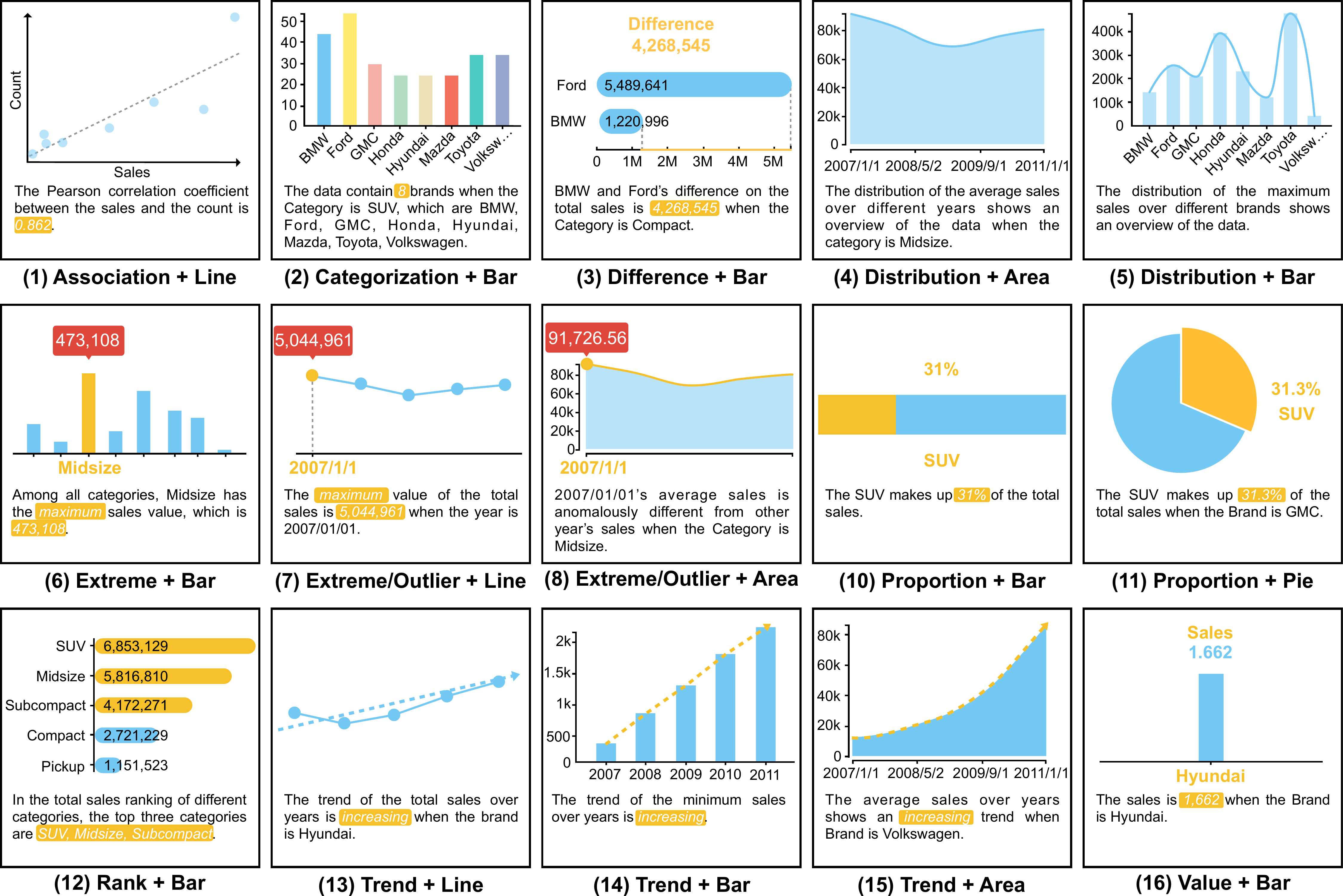}
\caption{The gallery of annotated charts designed for narrating the data content organized in 10 types of data facts. Each chart consists of a statistical diagram, annotations, and an automatically generated caption.}
\label{fig:charts}
\end{figure*}

A floating tool bar (Fig.~\ref{fig:table}(a3, b3)) is designed and placed at the bottom of both views, through which users can upload the data, ask a question via both voice and text input, enter the edit mode for editing the answers, and switch between the data and answer views.
We use the Web Speech API~\footnote{Web Speech API: \url{https://developer.mozilla.org/Web/API/Web_Speech_API}} to convert audio to text in the web browser.
Users can also enter the full-screen mode or covert the results into a PDF report by clicking the buttons in the up-right corner (Fig.~\ref{fig:table}(b4)). 

\subsection{Annotated Chart Library for Tabular Data}
\label{sec:chartlib}

In data storytelling, annotations in charts will help emphasize the information and avoid ambiguity~\cite{lee2015more}. Therefore, we design a library of annotated statistical charts~\footnote{The chart library will be released after the review} to display the data facts that answer the question with the goal of helping with the answer narration and interpretation. Our chart library consists of 5 types of basic statistic diagrams (bar chart, line chart, pie chart, area chart, and scatter plots) that are frequently used in data stories as summarized in ~\cite{shi2021autoclips}. To design the annotations, we further investigated a large number of relevant designs by exploring the design of the charts frequently used in over 200 data videos and over 1500 info-graphics. As a result, annotations involving text, colors, shapes, pointers, lines are designed for showing values, illustrating the trends and relations, highlighting anomalies and extremes, emphasizing differences and ranks, and differentiate categories and proportions. Applying annotations in the aforementioned 5 types of charts to represent different narrative semantics gives us 15 different annotated charts(Fig.~\ref{fig:charts}).
For example, we use dash-lines in a bar chart to emphasize difference but use trending lines in bar chart to illustrate trend, which result in two different annotated charts. 

Despite the above annotations, caption is another crucial component in each of the annotated charts. It usually describes the important data patterns in a nature language to help users quickly capture the information shown in the chart. To generate the caption, we adopt the sentence template for each type of fact introduced in~\cite{shi2020calliope}. Considering these templates may generate problematic descriptions that have grammar errors, our system enables a free editing function, through which users can easily edit the captions to fix the errors when necessary.


\subsection{Answer Facts Layout}
\label{sec:layout}
We display the answers to the input question in the form of a dashboard that could be easily displayed on a big screen to facilitate online discussions about the data in real-time. The annotated charts are arranged in order to facilitate reading and answer narration. In particular, we divide the screen into several regions and allocate to decomposed questions. Within each region, we arrange the charts in order according to their relevance to the corresponding questions, and place them one by one from left to right and top to bottom to facilitate reading. The size of each chart is determined by their relevance score to the question.

\section{Evaluation}
We estimate the design of the system via an example case, an ablation study, and a controlled~user~study. 

\subsection{Example Case}
We performed a case study by inviting an expert from a business school to explore and analyzing a marking dataset about car sales records. The dataset consists of four dimensions: year, sales value, model, and brand (Fig.~\ref{fig:teaser}). The user started with a complex fuzzy question (i.e., \textit{Type-II}) ``\textit{How is the sales?}". The system automatically decomposed the question into three relevant simple questions: ``which category has the highest sales?" (\textit{Q1.1}), ``what is the sales over years?" (\textit{Q1.2}), and  ``what is the overall value of the sales?" (\textit{Q1.3}), which are answered by three annotated charts showing the best selling model, the overall sales trend, and the total sales value as shown in Fig.~\ref{fig:teaser}.

\begin{figure}[!h]
\includegraphics[width=0.9\linewidth]{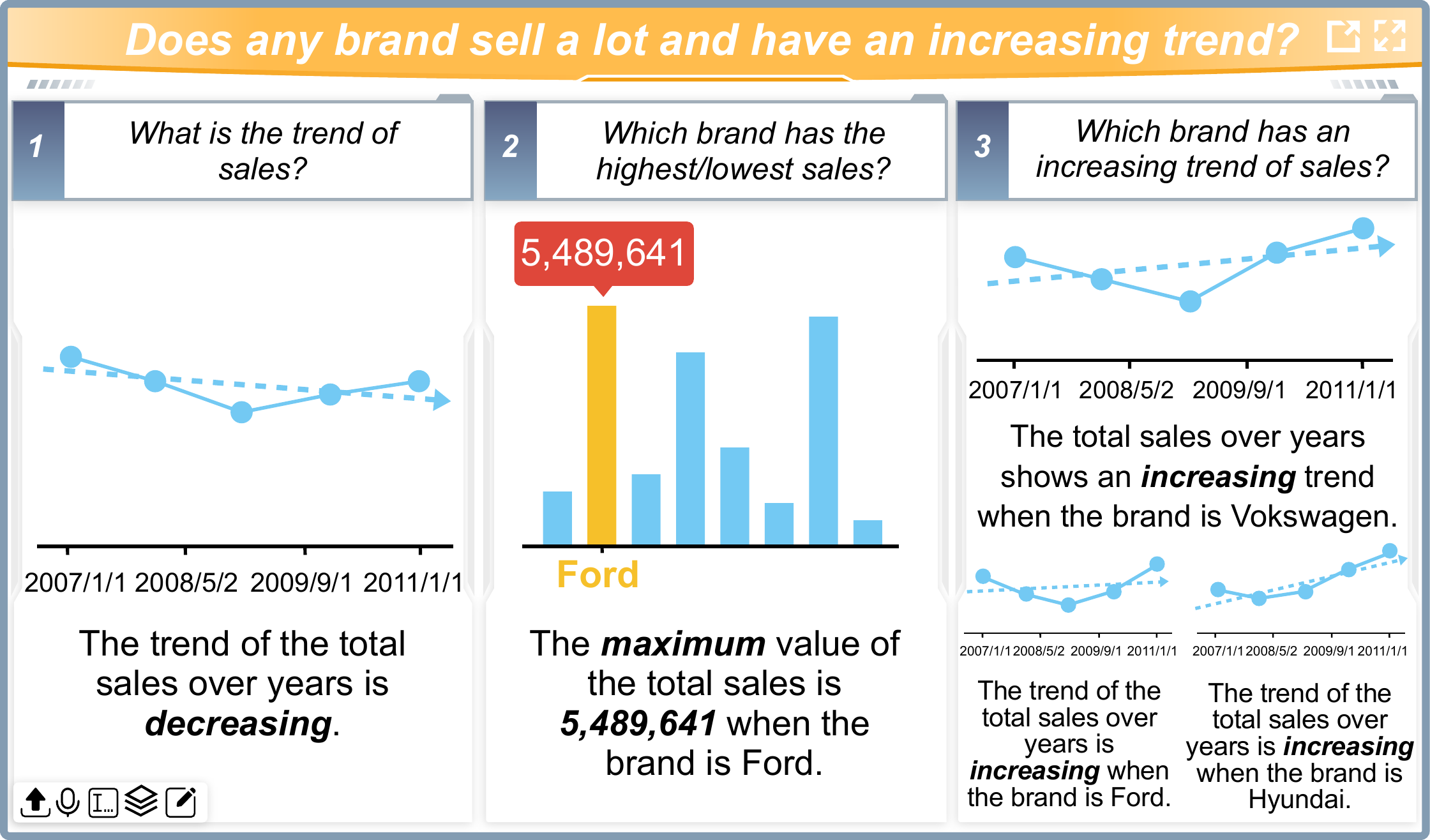}
  \caption{The user digs deeper into the Car Sales data by asking the system ``does any brand sell a lot and have an increasing trend?''. The system resolves the question into three relevant simple questions and presents corresponding annotated charts as answers.}
	\label{fig:dashboard}
\end{figure}

Having the above overview of the data, the user would like to dig deeper into the data. He asked ``does any brand sell a lot and have an increasing trend?" (i.e., \textit{Type-I}). The system resolves the question into three relevant simple questions: (1) ``what is the trend of sales?" (\textit{Q2.1}), (2)``which brand has the highest/lowest sales?" (\textit{Q2.2}), and (3) ``which brand has an increasing trend of sales?" (\textit{Q2.3}). The answers to these questions are shown in a group of charts as illustrated in Fig.~\ref{fig:dashboard}, which respectively illustrates the sales trend, showing the highest sales record among different brands and the sales trend of each brand. 

\subsection{Ablation Study}

We estimate the quality of the question decomposition results based on the testing corpus via two frequently used metrics in nature language processing, i.e., \textbf{BLEU}~\cite{papineni2002bleu} and \textbf{METEOR}~\cite{banerjee2005meteor}. These metrics are originally designed to estimate the quality of sentence translation. In particular, \textbf{BLEU} estimates the word-level translation precision based on the number of exactly matched words between the translated sentences and the ground-truth. \textbf{METEOR} computes a weighted F1-score to estimate the translation quality by comparing the translated sentences and ground-truth based on WordNet~\cite{miller1995wordnet}. These metrics are verified to be able to provide estimations that are consistent with humans' judgments~\cite{nema-khapra-2018-towards}. In our experiment, we use these matrices to estimate the decomposition quality by comparing the decomposed sub-questions to the corresponding targets in the testing~corpus.

We estimate the performance of the proposed decomposition model by comparing it to three simplified versions that respectively have (1) no copying mechanism, (2) no copying and attention mechanisms, and (3) no copying, attention, and question type classification components. All these models were trained based on the question decomposition corpus under the same parameters settings. In particular, the training set, validation set, and evaluation set respectively takes 80\%, 10\%, and 10\% of the corpus. We also involve a high-quality sentence rewriting technique as the baseline for comparison~\cite{xiao2020copy}. 

The evaluation results are summarized in Table~\ref{tab:performance}, which shows that (1) the sentence quality of our technique is equivalent to that of the high-quality sentence rewriting technique, and (2) our designs of the key components (question type classifier, attention, and copying mechanism) indeed improve the performance of the question decomposition model. 

\begin{table}[!h]
\begin{tabular}{p{8cm}p{1.2cm}p{1.8cm}}
\toprule
\textbf{Models} & \textbf{BLEU} & \textbf{METEOR} \\ \midrule
Sentence Rewriting Baseline & - & 22.12 \\
Decomposer & 23.88 & 24.02 \\ 
Decomposer + Classifier & 25.23 & 25.73 \\
Decomposer + Classifier + Attention & 25.56 & 26.09 \\ 
Decomposer + Classifier + Attention + Copying & \textbf{26.22} & \textbf{27.55} \\ 
\bottomrule
\end{tabular}
\vspace{0.5em}
\caption{Performance Evaluation of the Decomposition Model}
\label{tab:performance}
\end{table}

\subsection{User Study}
To estimate the usability of the system, we conducted a controlled within-subject study with 20 participants to make a comparison between \name and a baseline system developed based on NL4DV and vega-lite charts. The participants (13 female, 7 male, between 21 and 28 years old (M = 24.8, SD = 1.94)) are university students major in design and literature. They have limited knowledge about data analysis.

Two real datasets were used for the study. The first one describes 549 Amazon bestselling books (rows) from six dimensions (columns) including book title, rating, number of reviews, published year, price, and genre. The second dataset describes 275 car sales records from four dimensions including the sales value, brand, model, and the year. These two datasets were used in both the \name and the baseline system during the study in a counterbalanced order.

\begin{figure}[!h]
\centering 
\includegraphics[width=0.8\textwidth]{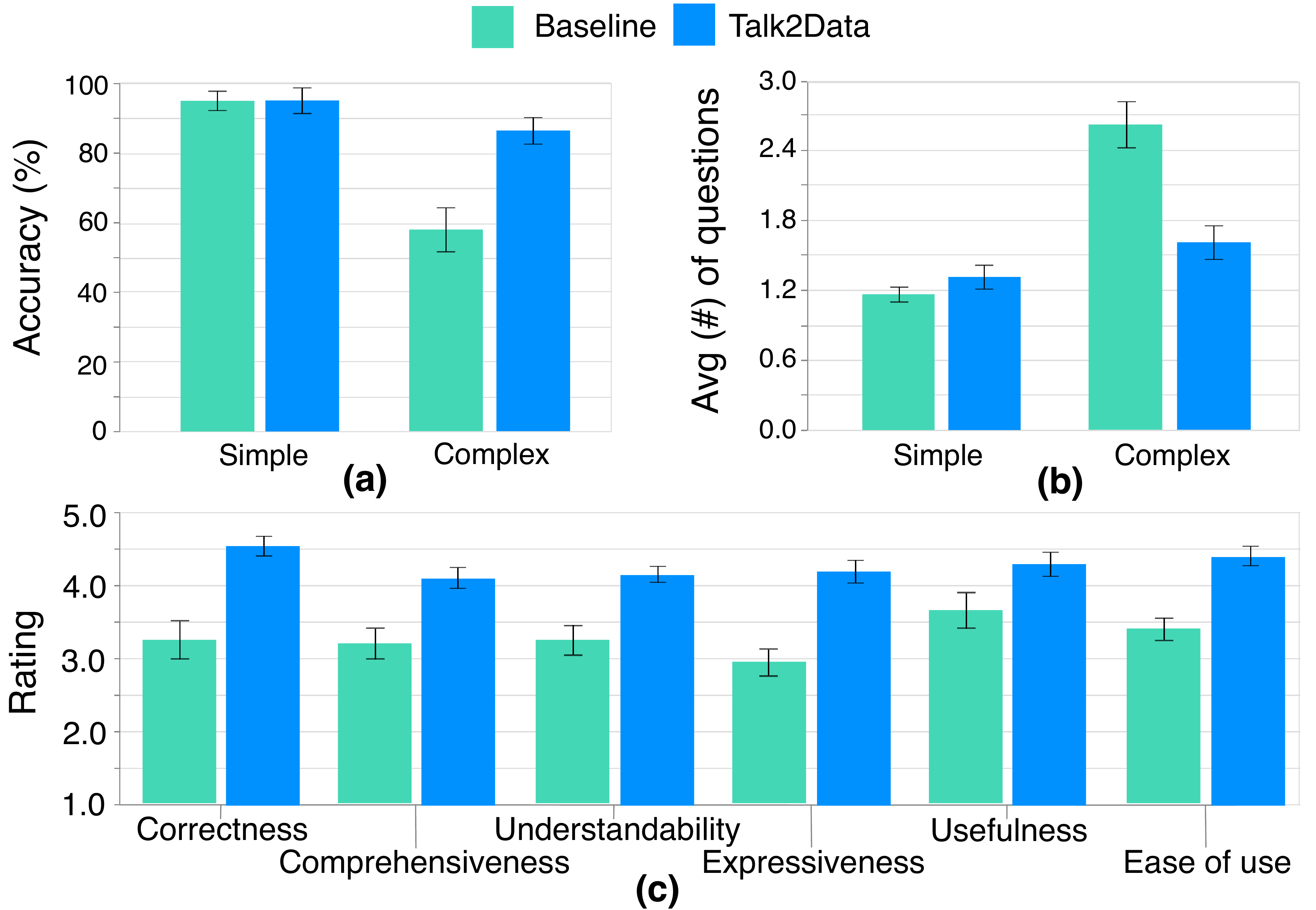}
\caption{The results of the user study: (a) the accuracy, (b) the averaged number of questions to finish each of the tasks, and (c) the ratings from different criteria based on a 5-point Likert scale, where 5 is the best and 1 is the worst.}
\label{fig:usertask}
\end{figure}

During the study, we first introduced the systems and let the participants to try it by their own. After the users were getting familiar with systems, they were asked to finish six tasks by asking relevant questions by their own. Three of these tasks were simple ones such as ``find the distribution of ratings over books" but the other three were complex ones such as ``find the most popular author". During the experiment, we recorded the number of questions asked by a user to finish each task in each system. This number were together with the accuracy to estimate their performance. Finally, the participants were also asked to finish a post-study questionnaire. The study results are reported as follows:

\textit{\underline{Accuracy.}} As shown in Fig.~\ref{fig:usertask}(a), \name (M=95\%, SD=0.16) and the baseline (M=95\%, SD=0.12) had a similar accuracy when finding answers to low level questions. However, our system (M=87.6\%, SD=0.17) significantly outperformed the baseline system (M=67.5\%, SD=0.28) in case of resolving complex questions based on the paired-t test (t(19) = 4.08, p \textless 0.01). 

\textit{\underline{Efficiency.}} We compared the averaged number of questions that users need to finish each of the tasks to demonstrate a system's efficiency. As shown in Fig.~\ref{fig:usertask}(b), when solving simple tasks, users explored a similar number of questions when using \name (M = 1.17,SD = 0.49) and the baseline (M = 1.32,SD = 0.76) system. However, for complex tasks, users obviously tend to ask more questions when using the baseline system (M = 2.63,SD = 1.53) comparing to that of \name (M = 1.62,SD = 1.11). The difference is significant regarding to the paired-t test (t(59) = 4.4, p \textless 0.01). 

\textit{\underline{Feedback.}} The results (Fig.~\ref{fig:usertask}(c)) of our post-study questionnaire showed that all the users preferred our system. Most of them mentioned ``\textit{\name is a useful tool}'', ``\textit{it can greatly save one's efforts when exploring the data}'', and ``\textit{the system is easy to use}". Many users also mentioned ``\textit{dividing a complex question into simple ones and answer them one by one is an intuitive and effective way to solve the problem}''. 

\section{Limitations and Future Work}

Here, we would like to report and discuss several limitations that was found during our system implementation and evaluation.

\underline{\textit{Scalability Issue.}} The current implementation of the prototype system still cannot handle large datasets that contain tens of thousands of data records, where the answer extraction algorithm is the primary bottleneck. It will be more difficult to find out accurate answers from a large dataset within a fixed period of time. There are several approaches that could be applied to address the issue, which will be our future work. First, using parallel searching algorithms~\cite{roosta2000parallel} will greatly improve the algorithm efficiency. Second, using a pre-trained model such as TaBERT~\cite{yin2020tabert}, to built a table-based Q\&A system, will also improve the system's performance. Although such a system doesn't exist yet, we believe it is a promising research direction.



\underline{\textit{Accuracy Issue.}} Although showing the relevant context is helpful for the answer interpretation, when mistake happens, the irrelevant charts could also be a distraction, which will affect users' judgments. We believe there are two methods that could be used to improve the accuracy of the system. First, we can employ knowledge bases such as WolframAlpha~\footnote{WolframAlpha: \url{https://www.wolframalpha.com/}} and knowledge graphs to guide the searching directions so that the answers could be more directly found without checking too many irrelevant candidates in the space. Second, again, training a QA system based on TaBERT~\cite{yin2020tabert} could also help improve the accuracy. 

\underline{\textit{Generalization Issue.}} Our training corpus is generated based on 26 tabular data that primarily contain marketing data records such as car sales values, and best selling books. As a result, our model could better handle complex questions in the marking domain, but may have a lower question decomposition quality when facing a question from other domains. To overcome the issue, more datasets in various domains should be collected and more questions should be prepared to train the model and improve the generalization of the system.
\section{Conclusion}

We present \name, a natural language interface for exploratory visual analysis that supports answering both simple and complex questions. It employs a deep-learning based question decomposition model to resolve a complex question into a series of relevant simple questions, and a search algorithm to extract the data facts that are most relevant to each of simple questions. To visualize the data facts, we designed a set of annotated and captioned visualization charts to support interpretation and narration. The proposed technique was evaluated via an ablation study and a controlled user study. The evaluation showed the power of the \name and revealed limitations and future work of the current system.

\begin{acks}
Nan Cao and Qing Chen are the corresponding authors. This work was supported in part by NSFC 62061136003, NSFC 6200070909, NSFC 62072338, and NSF Shanghai 23ZR1464700.
\end{acks}

\bibliographystyle{ACM-Reference-Format}
\bibliography{main}

\end{document}